\documentclass[prb,showpacs,twocolumn,amsmath,amssymb,superscriptaddress]{revtex4}
\usepackage{epsfig}
\usepackage{graphicx}
\usepackage{dcolumn}
\usepackage{bm}
\usepackage{amsthm}

\begin{document}
\title{Limited influence of diluted ferromagnetic dimers on Curie temperature in complex magnetic systems.}
\author{M.V.~Medvedev}
\affiliation{Institute of Electrophysics, Russian Academy of
Sciences-Ural Division, 620016 Yekaterinburg, Russia}
\author{I.A.~Nekrasov}
\affiliation{Institute of Electrophysics, Russian Academy of
Sciences-Ural Division, 620016 Yekaterinburg, Russia}

\date{\today}

\begin{abstract} 
In this work we investigate Ising and classical Heisenberg models for two and three dimensional
lattices in presence of diluted ferromagnetic dimers. For such models the Curie temperature as
a function of ratio of intra-dimer exchange coupling constant $I_A$ and other inter-site coupling
constants $I_B$ is calculated. In case dimer is treated exactly and environment within the mean-field approach
it was found that even for $I_A/I_B\to\infty$ $T_C$ remains finite. Similar analysis is proposed
for rhombohedral phase of intermetallic compound Gd$_2$Fe$_{17}$ where so-called Fe1-Fe1 ``dumbbell''
forms the diluted ferromagnetic dimer. It was shown that for such complex magnetic systems
$T_C$ is determined by all variety of exchange interactions and for the interval $0\leq I_A/I_B\leq\infty$
$T_C$ changes are not more than $\pm$10\%.

\end{abstract}


\maketitle 
\section{Introduction}

There is a number of magnetic compounds with complex crystal structure and
many magnetic ions per unit cell. Interatomic distances in such systems between
nearest exchange-coupled ionic pairs could be quite different by its value.
Thus the value of exchange coupling parameter for a pair with shortest
interatomic distance could be 2-3 times stronger than for a pair of atoms
with larger interatomic distance.

A typical example of such systems is  R$_2$Fe$_{17}$ class of intermetallic
compounds. Here shortest Fe1-Fe1 distance corresponds to Fe ions in a so called
``dumbbell'' positions (Wyckoff positions 4f for  hexagonal $Th_{2}Ni_{17}$-type
and 6c for rhombohedral $Th_{2}Zn_{17}$-type crystal structures, see for
details of magnetic structure~Ref.~\onlinecite{Kou98}).
Based on band structure calculations within the LSDA+U method
the exchange interaction parameter value for dumbbell pair
in the hexagonal phase Gd$_2$Fe$_{17}$
gives ferromagnetic exchange with  $I_{11}$(1)=238.8~K for a distance
$r_{11}$(1)=2.400\AA.\cite{Luk09} While for Fe3-Fe3 pairs in the 12j  Wyckoff positions
for a distance of $r_{33}$(1)=2.400~\AA corresponding exchange parameter value is found
to be  $I_{33}$(1)=80.4~K.\cite{Luk09}
Analogous calculations for rhombohedral phase of Gd$_2$Fe$_{17}$
for Fe1-Fe1 dumbbell pair gives ferromagnetic coupling with $I_{11}$(1)=287.5~K
($r_{11}$(1)=2.385~\AA).\cite{Luk09} For the next pair interatomic distance Fe2-Fe3
(9d and 18f Wychoff positions correspondingly) we obtained  $I_{23}$(1)=87.1~K
($r_{23}$(1)=2.423~\AA)\cite{Luk09}. Thus one can see
that coupling for dumbbell positions is 3 times stronger (in both phases)
than an exchange coupling for the next smallest pair interatomic Fe-Fe distance.

Let us note that these strong exchange bonds $I_{11}$(1) of Fe1-Fe1
dumbbell positions are rather well spatially separated one from each other
(rather diluted) and do not form any infinite magnetic cluster.
Moreover these strongly coupled dumbbell pairs one can consider as 
some ferromagnetic dimers embedded into a infinite magnetic cluster
formed by other weaker magnetic bonds.

To this end there rises a question: how strong influences presence of
such diluted ferromagnetic dimers on
a Curie temperature? Well could be that such dimers give dominating
contribution to the Curie temperature of a complex magnetic system
described above.

Here we solve a problem of influence of 
diluted ferromagnetic dimers on a Curie point. 
First for model lattices with Ising or classical
Haisenberg spins with different numbers of lattice sites
we vary exchange interaction strengths of the dimer from zero to infinity
while all other exchange interactions remains finite (see Sec.~\ref{s2}).
Then we investigate rhombohedral phase of Gd$_2$Fe$_{17}$ 
using the same approach (Sec.~\ref{s3}).
Finally we conclude our work with a summary (Sec.~\ref{conclusion}).

\section{
The Curie temperature for model lattices with diluted ferromagnetic dimers}
\label{s2}

\subsection{Ising spins case}
\label{isc}

\begin{figure}[t]
\begin{center}
\epsfxsize=8cm
\epsfbox{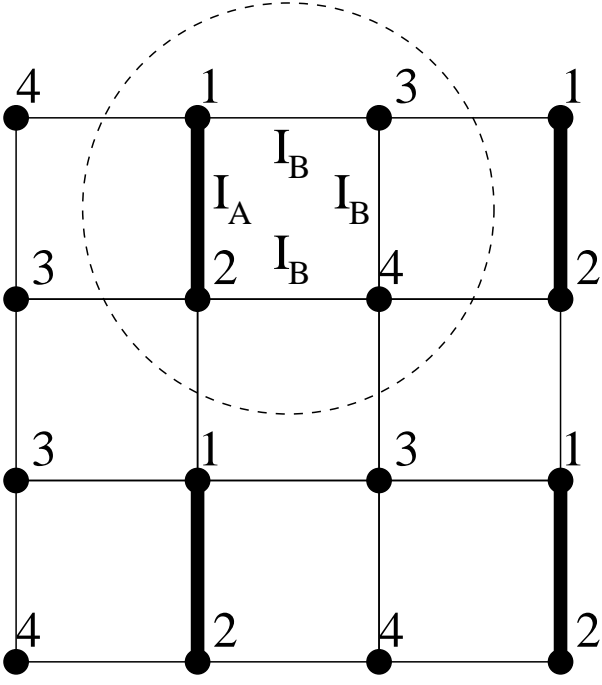}
\end{center}
\caption{
Square lattice with spins in the lattice sites with the period 2$a$ (magnetic unit cell surrounded by dashed circle).
Strong exchange bond $I_A$ --- fat black line, thick lines --- exchange bonds $I_B$.
Numbers 1,2,3,4 enumerate spins (magnetic sublattices) in the magnetic unit cell.}
\label{fig1}
\end{figure}

Let us consider square lattice with lattice parameter $a$. On each site
we have Ising spin $\tau_i=\pm 1$. We also specify that
spins  $\tau_1$ and  $\tau_2$ on a plaquette (selected on Fig.~\ref{fig1}
by dashed circle) form a dimer and are coupled by exchange integral $I_A>$0
while each other pair of
spins $\tau_1-\tau_3$,  $\tau_3-\tau_4$ and $\tau_4-\tau_2$
are coupled with $I_B>$0 and  $I_B \neq I_A$. Thus magnetic elementary cell contains
four magnetic atoms and the lattice has period 2$a$.

Using Weiss mean-field approximation it is easy to obtain four self-consistent
equations for observable spin value $\sigma_i=<\tau_i>$ ($i$=1,2,3,4) for each magnetic sublattice.

\begin{eqnarray}
\label{eq1}
\sigma_1 & = & th\{ \beta [I_A\sigma_2 + I_B(\sigma_2 + 2\sigma_3)] \} \nonumber \\
\sigma_2 & = & th\{ \beta [I_A\sigma_1 + I_B(\sigma_1 + 2\sigma_4)] \} \nonumber \\
\sigma_3 & = & th\{ \beta I_B(2\sigma_1 + 2\sigma_4) \} \nonumber \\
\sigma_4 & = & th\{ \beta I_B(2\sigma_2 + 2\sigma_3) \}
\end{eqnarray}
with $\beta=1/k_BT$.

If $T \rightarrow T_C$ and $\sigma_1,\sigma_2,\sigma_3,\sigma_4 \rightarrow 0$ 
Eq.~\ref{eq1} reduces to linear equations determining $T_C$

\begin{eqnarray}
\label{eq2}
\sigma_1 & = & \beta_C [(I_A+I_B)\sigma_2 + 2I_B\sigma_3] \nonumber \\
\sigma_2 & = & \beta_C [(I_A+I_B)\sigma_1 + 2I_B\sigma_4] \nonumber \\
\sigma_3 & = & \beta_C I_B(2\sigma_1 + 2\sigma_4)  \nonumber \\
\sigma_4 & = & \beta_C I_B(2\sigma_2 + 2\sigma_3).
\end{eqnarray}

For ferromagnetic ordering in elementary magnetic cell 
shown on Fig.~\ref{fig1} follows relations $\sigma_1=\sigma_2$ and $\sigma_3=\sigma_4$.
Then 

\begin{equation}
\label{eq3}
k_BT_C = \frac{1}{2}\Bigl(I_A+3I_B+\sqrt{I_A^2-2I_AI_B+17I_B^2}\Bigr)
\end{equation}

If $I_A=0$ it comes out from Eq.~(\ref{eq3}) $k_BT_C(I_A=0)=\frac{1}{2}(3+\sqrt{17})I_B=3.562I_B$ and
for $I_A=I_B$ one can find  $k_BT_C(I_A=I_B)=4I_B$. However in limiting case
$I_A\rightarrow\infty$ Eq.~(\ref{eq3}) leads to $k_BT_C\rightarrow\infty$.

It is obvious that the above mentioned result is not correct from physical point of view.
In case $I_A \gg I_B$ strong ferromagnetic exchange coupling $I_A$ within a dimer
preserves parallel alignment of spins $\tau_1$ and $\tau_2$.
While decoupling of weak exchange bonds $I_B$ 
(which form infinite ferromagnetic cluster)
is responsible for destruction of ferromagnetic order.
Thus even for $I_A\rightarrow\infty$ Curie temperature $T_C$ should be finite and proportional
to $I_B$.

The origin of such paradox is applied mean-field approximation 
where strong fluctuations of exchange energy
because of spin flips are absent.
The fluctuations become anomalously strong if we deal with groups
of spins strongly coupled with each other.
To this end one should use approach proposed by
Scalapino, Imry and Pincus\cite{Scal75} for quasi one dimensional
magnetic systems with two significantly different exchange constants
$I_A$ and $I_B$. The main idea is to treat strongly coupled spins exactly
while weak exchange interactions are treated within the mean-field approximation.

For this purpose let us write down a Hamiltonian of such two spins cluster (dimer)
in the elementary magnetic cell shown on Fig.~\ref{fig1}
\begin{eqnarray}
\label{eq4}
{\bf H}_{cl} = -I_A\tau_1 \tau_2 & - & I_B (\sigma_2 + 2\sigma_3) \tau_1  - \nonumber \\
                                 & - & I_B (\sigma_1 + 2\sigma_4) \tau_2.
\end{eqnarray}

Here interaction between spins $\tau_1$ and $\tau_2$ in the dimer which are
coupled with a large exchange constant $I_A$ is kept in the exact form and
interaction of the same spins $\tau_1$ and $\tau_2$ with others via
small $I_B$ is treated within the mean-field approximation.

Partition function for the dimer in the environment is
\begin{eqnarray}
\label{eq5}
Z_{cl}& = & 2\Bigl\{ e^{ \beta I_A} ch [\beta I_B (\sigma_1 + \sigma_2 + 2\sigma_3 + 2\sigma_4)] + \nonumber \\
             & + &          e^{-\beta I_A} ch [\beta I_B (\sigma_1 - \sigma_2 - 2\sigma_3 + 2\sigma_4)] \Bigr\}. 
\end{eqnarray}

Then for average values of spins $\sigma_1=<\tau_1>$ and  $\sigma_2=<\tau_2>$ one can
derive self-consistent set of equations

\begin{eqnarray}
\label{eq6}
\sigma_1 & = & \frac{2}{Z_{cl}} \Bigl\{ e^{ \beta I_A} sh [\beta I_B (\sigma_1 + \sigma_2 + 2\sigma_3 + 2\sigma_4)] - \nonumber \\
             & - &          e^{-\beta I_A} sh [\beta I_B (\sigma_1 - \sigma_2 - 2\sigma_3 + 2\sigma_4)] \Bigr\},
\end{eqnarray}
\begin{eqnarray}
\label{eq7}
\sigma_2 & = & \frac{2}{Z_{cl}} \Bigl\{ e^{ \beta I_A} sh [\beta I_B (\sigma_1 + \sigma_2 + 2\sigma_3 + 2\sigma_4)] + \nonumber \\
             & + &          e^{-\beta I_A} sh [\beta I_B (\sigma_1 - \sigma_2 - 2\sigma_3 + 2\sigma_4)] \Bigr\}.
\end{eqnarray}
For average values of spins  $\sigma_3=<\tau_3>$ and  $\sigma_4=<\tau_4>$ 
weakly coupled to environment via $I_B$ still valid last two equations of Eqs.~(\ref{eq1}). 

It is remarkable that for the limit $I_A\rightarrow\infty$ immediately follows
\begin{equation}
\label{eq8}
\sigma_1 = \sigma_2 = th [\beta I_B (\sigma_1 + \sigma_2 + 2\sigma_3 + 2\sigma_4)].
\end{equation}

It means that for infinitely strong exchange coupling between spins  $\sigma_1=<\tau_1>$ and  $\sigma_2=<\tau_2>$
(in case the spins are strictly parallel to each other)
environment mean-field $I_B(\sigma_1+2\sigma_4)$ acting on spin $\tau_2$ which belongs to the dimer
is transferred onto spin $\tau_1$ and thus is added to Weiss field $I_B(\sigma_2+2\sigma_3)$.
Similar ``transfer'' of Weiss field from spin  $\tau_1$ happens to spin $\tau_2$.

If we turn back to finite values of $I_A$, considering $\sigma_1=\sigma_2$ and  $\sigma_3=\sigma_4$
because of symmetry of magnetic surrounding, one can obtain two self-consisted equations 
\begin{eqnarray}
\label{eq9}
\sigma_1 & = & \frac{sh[\beta I_B(2\sigma_1+4\sigma_3)]}{ch[\beta I_B(2\sigma_1+4\sigma_3)]+e^{-2 \beta I_A}}, \nonumber \\ 
\sigma_3 & = & th [2\beta I_B(\sigma_1+\sigma_3)].
\end{eqnarray}

In case of $T\rightarrow T_C$ one can linearize Eqs.~(\ref{eq9}) for
$\sigma_1,\sigma_3\rightarrow0$ and get uniform linear set of equations

\begin{eqnarray}
\label{eq10}
\sigma_1 & = & \beta_C I_B[1+th(\beta_C I_A)](\sigma_1+2\sigma_3), \nonumber \\ 
\sigma_3 & = & 2\beta_C I_B(\sigma_1+\sigma_3).
\end{eqnarray}

Since Eqs.~(\ref{eq10}) are valid only for $I_A\gg I_B$, then
for $I_A=10I_B$ one can find $k_BT_C(I_A=10I_B)=4.792I_B$ and for
$I_A\rightarrow\infty$ correspondingly $k_BT_C(I_A\rightarrow\infty)=2(1+\sqrt{2})I_B=4.828I_B$.

To summarize this subsection we have investigated a square lattice model
with four spins in the elementary magnetic cell
and one strong exchange bond $I_A$ out of four.
Varying the selected exchange coupling constant in the interval $0<I_A<\infty$ we obtained
variation of the Curie point from  $k_BT_C(I_A=0)=3.562I_B$ upto $k_BT_C(I_A\rightarrow\infty)=4.828I_B$.

\subsection{Elementary magnetic cell finite size effect on $T_C$}
\label{fse}

\begin{figure}[t]
\begin{center}
\epsfxsize=8cm
\epsfbox{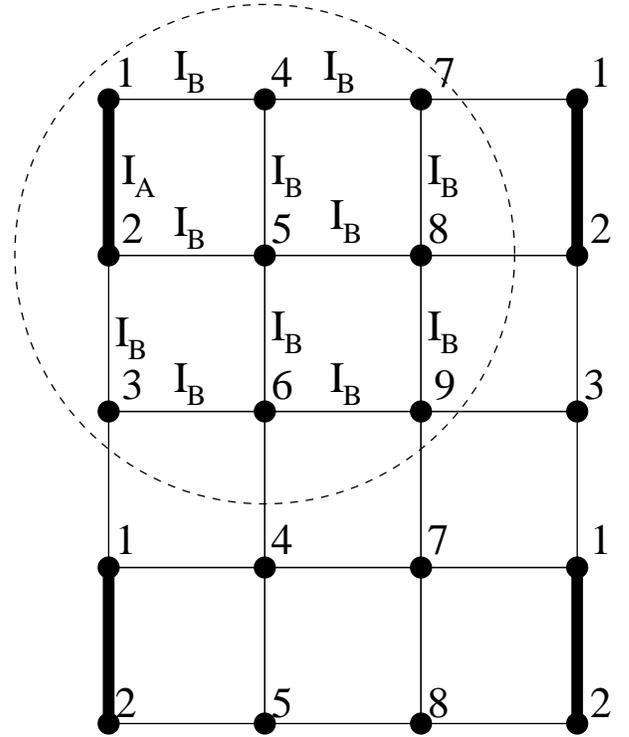}
\end{center}
\caption{
The same as Fig.~1, but for the period 3$a$ (9 magnetic sublattices).}
\label{fig2}
\end{figure}

Now one traces how an increase of number of magnetic sites in the elementary magnetic cell
changes Curie temperature in presense of the dimer.
In case of square lattice with the periodicity of magnetic elementary cell equal to 3$a$ containing
nine magnetic sites (see Fig.~\ref{fig2}) magnetic structure is described by nine magnetic
sublattices. However using symmetry of magnetic environment one can find following
relations between different sublattice spin values
$\sigma_1=\sigma_2$, $\sigma_4=\sigma_5=\sigma_7=\sigma_8$ and $\sigma_6=\sigma_9$.
Thus if $0\leq I_A\lesssim I_B$ employing mean-field approximation for 
 $T\rightarrow T_C$ one can obtain four linear equations
\begin{eqnarray}
\label{eq12}
\sigma_1 & = & \beta_C [I_A\sigma_1 + I_B(\sigma_3+2\sigma_4)], \nonumber \\
\sigma_3 & = & \beta_C [I_B(2\sigma_1 + 2\sigma_6)], \nonumber \\
\sigma_4 & = & \beta_C [I_B(\sigma_1 + 2\sigma_4+\sigma_6)],  \nonumber \\
\sigma_6 & = & \beta_C [I_B(\sigma_3 + 2\sigma_4+\sigma_6)].
\end{eqnarray}
By solving fourth order determinant one can get $k_BT_C(I_A=0)=3.820I_B$.

In case of strong exchange bond $I_A\gg I_B$ it is necessary to write
corresponding Hamiltonian similar to Eq.~(\ref{eq4})
using relations  $\sigma_4=\sigma_5=\sigma_7=\sigma_8$
\begin{eqnarray}
\label{eq13}
{\bf H}_{cl}(1,2) = -I_A\tau_1 \tau_2 & - & I_B (\sigma_3 + \sigma_4 + \sigma_7) \tau_1 - \nonumber \\
                                      & - & I_B (\sigma_3 + \sigma_5 + \sigma_8) \tau_2 = \nonumber \\
                  = -I_A\tau_1 \tau_2 & - & I_B (\sigma_3 + 2\sigma_4)(\tau_1 + \tau_2).
\end{eqnarray}
Then average value $\sigma_1=<\tau_1>$ is
\begin{equation}
\label{eq14}
\sigma_1  =  \frac{sh[\beta I_B(2\sigma_3+4\sigma_4)]}{ch[\beta I_B(2\sigma_3+4\sigma_4)]+e^{-2 \beta I_A}}.
\end{equation}
In the vicinity of $T_C$ Eq.~(\ref{eq14}) is linearized for small values
$\sigma_3,\sigma_4\rightarrow 0$ and gives 
\begin{equation}
\label{eq15}
\sigma_1  =  \beta_C I_B[1+th(\beta_C I_A)](\sigma_3+2\sigma_4),
\end{equation}
and should be used as a first equation in the Eqs.~(\ref{eq12}).
Then one can calculate
$k_BT_C(I_A=10I_B)=4.331I_B$ and $k_BT_C(I_A\rightarrow\infty)=4.340I_B$.

\begin{figure}[t]
\begin{center}
\epsfxsize=8cm
\epsfbox{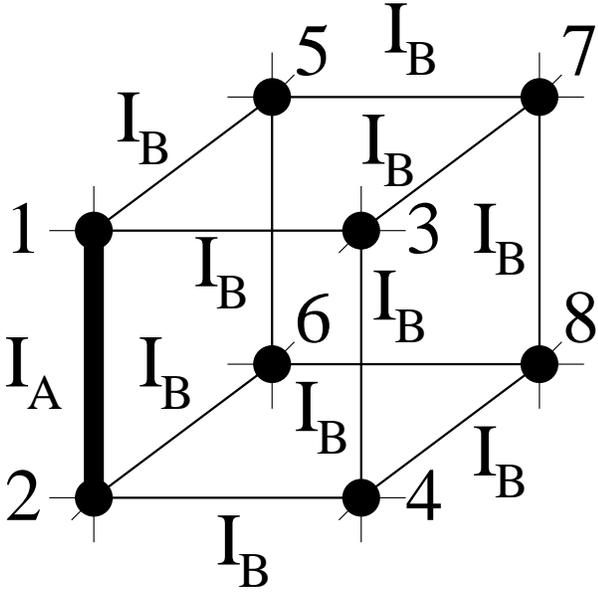}
\end{center}
\caption{
The same as Fig.~1, but for the three dimensional cubic lattice (8 magnetic sublattices).}
\label{fig3}
\end{figure}

Similar calculations were done by us also for the case of two dimensional lattice with period 4$a$ 
as well as for three dimensional models with elementary magnetic cells $(2a\times 2a\times 2a)$
shown on Fig.~\ref{fig3} (8 magnetic sublattices),  $(3a\times 3a\times 2a)$ (18 magnetic sublattices)
and $(3a\times 3a\times 3a)$ (27 magnetic sublattices). Corresponding results are
collected in Tabble~\ref{tab1}.
Is is clearly seen that under increase of number of magnetic sites
in the elementary magnetic cell and corresponding increase of number of
weak magnetic bonds influence of the dimer
on $T_C$ goes down and becomes insignificant even in the limit 
$I_A\rightarrow\infty$.

\begin{figure}[t]
\begin{center}
\epsfxsize=8cm
\epsfbox{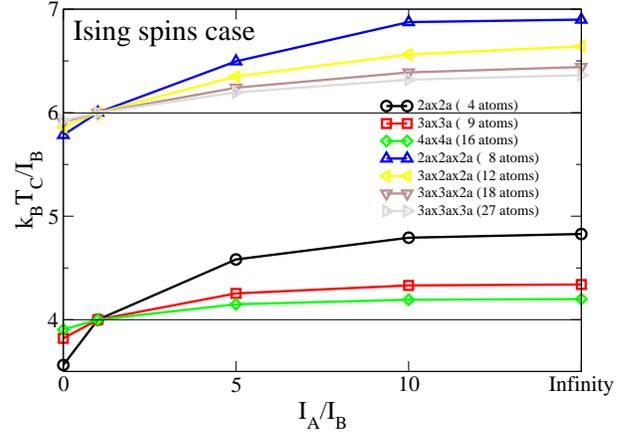}
\end{center}
\caption{Dimensionless Curie temperature versus selected exchange bond strength to other exchange bond strengths ratio $I_A/I_B$
for two and three dimensional cubic lattices of Ising spins with different size of elementary magnetic cells.}
\label{fig4}
\end{figure}

Roughly speaking infinitely strong exchange bond $I_A$ (in case spins are
strictly parallel in the dimer) produces an effective spin with doubled
value of magnetic moment. At the same time it decreases number of magnetic sites
by one. After all such dimer contributes rather weak into free energy of the system
in comparison with other spins.
The same data is displayed on Fig.~\ref{fig4}. Here one can see a tendency for the
$T_C$ to lesser and lesser deviate from $T_C(I_A=I_B)$ for larger number of atoms
in the elementary magnetic cell.

\begin{table*}
\caption{Curie temperature value for different magnetic elementary cells with Ising spins
with selected strong exchange bond $I_A$ as a function of ratio $I_A/I_B$.
$T_C$ is given in units $I_B$.}
\begin{ruledtabular}
\begin{tabular}{|c|c|c|c|c|c|c|}
Magnetic               & Magnetic  &          &              &               &                &                        \\
elementary             & atoms     & $I_A$=0  & $I_A$=$I_B$  & $I_A$=5$I_B$  & $I_A$=10$I_B$  & $I_A\rightarrow\infty$ \\
cell                   & number    &          &              &               &                &                        \\
\hline
$2a\times 2a$          & 4         & 3.562    & 4            & 4.582         & 4.792          & 4.828                 \\
\hline
$3a\times 3a$          & 9         & 3.820    & 4            & 4.254         & 4.331          & 4.340                 \\
\hline
$4a\times 4a$          & 16        & 3.905    & 4            & 4.149         & 4.193          & 4.199                 \\
\hline
\hline
$2a\times 2a\times 2a$ & 8         & 5.785    & 6            & 6.496         & 6.785          & 6.899                 \\
\hline
$3a\times 2a\times 2a$ & 12        & 5.862    & 6            & 6.350         & 6.562          & 6.641                 \\
\hline
$3a\times 3a\times 2a$ & 18        & 5.910    & 6            & 6.241         & 6.389          & 6.441                 \\
\hline
$3a\times 3a\times 3a$ & 27        & 5.929    & 6            & 6.196         & 6.319          & 6.362
\end{tabular}
\end{ruledtabular}
\label{tab1}
\end{table*}

\subsection{
Classical Heisenberg spins case}
\label{chs}

In our days there is a way to compute exchange interaction parameters
between different sites of Heisenberg model with classical spins
proposed by Lichtenstein $et~al$. in Ref.~\onlinecite{Licht87}.
In this approach exchange parameters are calculated as a second derivative
of total energy with respect to small angles of magnetic moments deviation from
collinear magnetic configuration. Thus one should understand
how Curie temperature of ferromagnetic lattice of classical Heisenberg
spins depends on presence of diluted magnetic dimers.
Here we present analysis similar to Sec.~\ref{isc}.

At the beginning we consider a lattice shown on Fig.~\ref{fig1}
with classical spin vectors ${\vec S}_i=S(sin\theta_i cos\phi_i, sin\theta_i sin\phi_i, cos\theta_i)$.
Now we calculate average values of spin vectors for ferromagnetic case
assuming that spontaneous magnetic moment is directed along $z$-axis.
It means that $<S^x_{i}>=<S^y_{i}>=0$ and $m_i\equiv <S^z_{i}>=S<cos\theta_{i}>\neq0$.
Within the mean-field approximation for spins average value on sites of each of four
magnetic sublattices one can get following equations:

\begin{eqnarray}
\label{eq17}
m_1 & = & S{\cal L} ( \beta S [I_A m_2 + I_B(m_2 + 2m_3)] ),  \nonumber \\
m_2 & = & S{\cal L} ( \beta S [I_A m_1 + I_B(m_1 + 2m_4)] ),  \nonumber \\
m_3 & = & S{\cal L} ( \beta S I_B[2m_1 + 2m_4] ),             \nonumber \\
m_4 & = & S{\cal L} ( \beta S I_B[2m_2 + 2m_3] ).
\end{eqnarray}
where ${\cal L}(x)=cth(x)-\frac{1}{x}$ -- Langevin function.
For $T\rightarrow T_C$  ${\cal L}(x)\approx \frac{x}{3},~x\ll 1$
(compare with Eq.~\ref{eq2}):
\begin{eqnarray}
\label{eq18}
m_1 & = & \beta_C \frac{S^2}{3} [(I_A+I_B)m_2 + 2I_Bm_3], \nonumber \\
m_2 & = & \beta_C \frac{S^2}{3} [(I_A+I_B)m_1 + 2I_Bm_4], \nonumber \\
m_3 & = & \beta_C \frac{S^2}{3} I_B(2m_1 + 2m_4),         \nonumber \\
m_4 & = & \beta_C \frac{S^2}{3} I_B(2m_2 + 2m_3).
\end{eqnarray}
Exploiting symmetry relations $m_1=m_2$ and $ m_3=m_4$ one can end up with
equation on $T_C$
\begin{equation}
\label{eq19}
k_BT_C = \frac{S^2}{3} \frac{1}{2}\Bigl(I_A+3I_B+\sqrt{I_A^2-2I_AI_B+17I_B^2}\Bigr),
\end{equation}
which differs from Eq.~(\ref{eq3}) only with coefficient $\frac{S^2}{3}$ on the left side.
Similar to Ising case simply mean-field treatment gives incorrect results,
for the case  $\frac{I_A}{I_B}\rightarrow\infty$
namely, $T_C\rightarrow\infty$.

Let us employ once more the approach of Ref.~\onlinecite{Scal75}, where
for the case $I_A\gg I_B$ for a dimer we write down a Hamiltonian with
exchange interaction in the dimer is taken exactly.
The exchange coupling constant is $I_A$ and spin vectors $\vec S_1$ $\vec S_2$.
Coupling of these two spins to other ones via weak exchange integral $I_B$
we consider within mean-field approximation:
\begin{equation}
\label{eq20}
{\bf H}_{cl}(1,2) = -I_A\vec S_1 \vec S_2 - h_1 (S^z_1+S^z_2),
\end{equation}
here Weiss field is $h_1=I_B(m_1+2m_3)$.
Corresponding partition function is
\begin{eqnarray}
\label{eq21}
Z_{cl}(1,2) & = & \!\!\int \!\!d\Omega_1\!\!\int \!\!d\Omega_2 exp\{\beta [I_A\vec S_1 \vec S_2 + h_1 (S^z_1+S^z_2)]\} = \nonumber \\
            & = & 4\pi\int d\Omega_1 exp(\beta S h_1 cos\theta_1)\times \nonumber \\
            & \times & \frac{sh(\beta S\sqrt{I_A^2S^2+2I_ASh_1cos\theta_1 +h_1^2})}
                            {\beta S\sqrt{I_A^2S^2+2I_ASh_1cos\theta_1 +h_1^2}},
\end{eqnarray}
where $d\Omega_i=sin\theta_id\theta_id\phi_i$ is solid angle element with
$0<\theta_i<\pi,~0<\phi_i<2\pi$.

\begin{figure}[t]
\begin{center}
\epsfxsize=8cm
\epsfbox{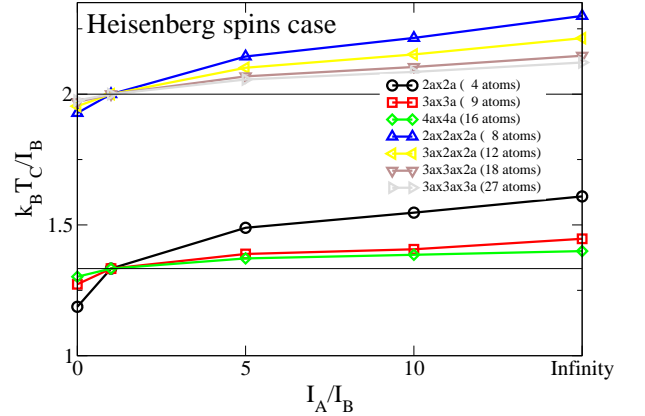}
\end{center}
\caption{The same as Fig.~4, but for classical Heisenberg spins.}
\label{fig5}
\end{figure}

Since to get $T_C$ one needs to linearize equations for $m_1$ and $m_3$ (that is $h_1$).
Therefore we calculate $Z_{cl}(1,2)$ up to the order $h_1^2$.
\begin{eqnarray}
\label{eq22}
Z_{cl}(1,2) & = & (4\pi)^2\frac{sh(\beta I_A S^2)}{\beta I_A S^2} \bigl\{ 1 + \nonumber \\
            & + & \frac{1}{3}[1+{\cal L}(\beta I_A S^2)](\beta S h_1)^2+...~\bigr\}.
\end{eqnarray}
Then average value of spin $m_1$ on site 1 in the elementary magnetic cell is
\begin{eqnarray}
m_1 & = & \frac{1}{2}(m_1+m_2)=\frac{1}{2}(<S_1^z>+<S_2^z>)= \nonumber \\
    & = & \frac{1}{2} \frac{1}{Z_{cl}(1,2)} \frac{\partial}{\partial(\beta h_1)}Z_{cl}(1,2) \nonumber.
\end{eqnarray}
After differentiating of Eq.~(\ref{eq22}) and following expanding of self-consistent
equation for $m_3$ around small values of $m_1$ and $m_3$ we come to two equations
\begin{eqnarray}
\label{eq23}
m_1 & = &  \beta_C I_B \frac{S^2}{3} [1+{\cal L}(\beta_C I_A S^2)] (m_1 + 2m_3), \nonumber \\
m_3 & = & 2\beta_C I_B \frac{S^2}{3} (m_1 + m_3).
\end{eqnarray}
Last equations for $T_C$ coincides with Eqs.~(\ref{eq10}) for Ising spins upto substitutions
$I_B\frac{S^2}{3}\rightarrow I_B$ and ${\cal L}(\beta_C I_A S^2)\rightarrow th(\beta_C I_A)$.
It gives us a way to transform corresponding equations determining $T_C$ from Sec.~\ref{isc} to
the case of classical Hesenberg spins in frame of the same model approach.
Let us also mention that limits $\lim_{I_A\to\infty}{\cal L}(\beta_C I_A S^2)=1$ and
$\lim_{I_A\to\infty}th(\beta_C I_A)=1$ coincide in magnitude, while for
finite values of $I_A$ values of $th(\beta_C I_A)$ and ${\cal L}(\beta_C I_A S^2)$
are slightly different (if $S=1$ is assumed). In the Table~\ref{tab2} 
values of $T_C$ are listed for the case of $S=1$ for different strength of
exchange coupling within dimer $I_A$. From Fig.~\ref{fig5} one can see that
values of $T_C$ vary not more than 10\% from $T_C(I_A=I_B)$ when $0<I_A<\infty$.
Again with increase of number of atoms in elementary magnetic cell $T_C$ is getting
closer and closer to $T_C(I_A=I_B)$.

\begin{table*}
\caption{Curie temperature value for different magnetic elementary cells with classical Heisenberg spins
with selected strong exchange bond $I_A$ as a function of ratio $I_A/I_B$ for $S=1$.
$T_C$ is given in units $I_B$.}
\begin{ruledtabular}
\begin{tabular}{|c|c|c|c|c|c|c|}
Magnetic               & Magnetic  &          &              &               &                &                        \\
elementary             & atoms     & $I_A$=0  & $I_A$=$I_B$  & $I_A$=5$I_B$  & $I_A$=10$I_B$  & $I_A\rightarrow\infty$ \\
cell                   & number    &          &              &               &                &                        \\
\hline
$2a\times 2a$          & 4         & 1.187    & 1.333        & 1.489         & 1.547          & 1.609                 \\
\hline
$3a\times 3a$          & 9         & 1.273    & 1.333        & 1.389         & 1.407          & 1.447                 \\
\hline
$4a\times 4a$          & 16        & 1.302    & 1.333        & 1.372         & 1.386          & 1.400                 \\
\hline
\hline
$2a\times 2a\times 2a$ & 8         & 1.928    & 2            & 2.144         & 2.215          & 2.299                 \\
\hline
$3a\times 2a\times 2a$ & 12        & 1.954    & 2            & 2.101         & 2.152          & 2.214                 \\
\hline
$3a\times 3a\times 2a$ & 18        & 1.970    & 2            & 2.068         & 2.104          & 2.147                 \\
\hline
$3a\times 3a\times 3a$ & 27        & 1.976    & 2            & 2.056         & 2.085          & 2.121
\end{tabular}
\end{ruledtabular}
\label{tab2}
\end{table*}

\section{Influence of dumbbell exchange strength on $T_C$ of the
rhombohedral phase of Gd$_2$Fe$_{17}$}
\label{s3}

Here we examine influence of strongly exchange coupled dumbbell Fe atoms
on  $T_C$ of the rhombohedral phase of Gd$_2$Fe$_{17}$.
In this compound there are 17 magnetic (Fe) atoms in the elementary 
magnetic cell, which occupies four types of nonequivalent crystallographic positions.
Correspondingly they have different surroundings of neighboring Fe atoms.
Namely, for $Th_2Zn_{17}$-type rhombohedral structure there are
2 Fe1 atoms in 6c Wychoff positions (dumbbell positions),
3 Fe2 atoms (9d), 6 Fe3 atoms (18f)  and 6 Fe4 atoms (18h).
Local magnetic moments of different classes of Fe atoms
are a bit different from each other:
$\mu_{Fe1}=2.19\mu_B$, $\mu_{Fe2}=2.26\mu_B$, $\mu_{Fe3}=2.17\mu_B$
and $\mu_{Fe4}=2.31\mu_B$.\cite{Luk09}
Exchange interaction parameters for the first coordination sphere of
different Fe atom classes calculated within the LSDA+U
approach\cite{Anisimov97} in Ref.~\onlinecite{Luk09} are
presented in Table~\ref{tab3}.
\begin{table*}
\caption{Parameters of exchange in the rhombohedral structure 
of Gd$_2$Fe$_{17}$ for the ions of the first coordination sphere.}
\begin{ruledtabular}
\begin{tabular}{ccccc}
N & Exchange (K) & Distance (\AA) & Number of neighbors & Type\\
\hline
1 & $I_{11}$(1)=287.5 & $r_{11}$(1)=2.385 & $z_{11}$(1)=1 & 
Fe1 (dumbbell) - Fe1 (dumbbell)\\
2 & $I_{44}$(1)=182.2 & $r_{44}$(1)=2.490 & $z_{44}$(1)=2 & 
Fe4 (corrugated plane) - Fe4 (corrugated plane)\\
3 & $I_{34}$(1)=125.9 & $r_{34}$(1)=2.551 & $z_{34}$(1)=$z_{43}$(1)=2 & 
Fe3 - Fe4 (corrugated plane)\\
4 & $I_{24}$(1)=121.0 & $r_{24}$(1)=2.448 & $z_{24}$(1)=4, $z_{42}$(1)=2 & 
Fe2 (corrugated layer) - Fe4 (corrugated layer)\\
5 & $I_{34}$(2)=105.7 & $r_{34}$(2)=2.613 & $z_{34}$(2)=$z_{43}$(2)=2 & 
Fe3 - Fe4 (corrugated layer)\\
6 & $I_{14}$(1)=88.2 & $r_{14}$(1)=2.639 & $z_{14}$(1)=3, $z_{41}$(1)=1 & 
Fe1 (dumbbell) - Fe4 (corrugated layer)\\
7 & $I_{23}$(1)=87.1 & $r_{23}$(1)=2.423 & $z_{23}$(1)=4, $z_{32}$(1)=2 & 
Fe3 (dumbbell) - Fe2 (corrugated layer)\\
8 & $I_{12}$(1)=83.6 & $r_{12}$(1)=2.602 & $z_{12}$(1)=3, $z_{21}$(1)=2 & 
Fe1 (dumbbell) - Fe2 (corrugated layer)\\
9& $I_{13}$(1)=74.1 & $r_{13}$(1)=2.740 & $z_{13}$(1)=6, $z_{31}$(1)=2 & 
Fe1 (dumbbell) - Fe3\\
10& $I_{33}$(1)=--36.5 & $r_{33}$(1)=2.466 & $z_{33}$(1)=2 & 
Fe3 - Fe3\\
\end{tabular}
\end{ruledtabular}
\label{tab3}
\end{table*}

It was found that largest parameter of exchange interaction $I_{11}(1)=287.5$~K
couples two Fe1 atoms in the dumbbell position with a distance inbetween $r_{11}$=2.385~\AA.
Next largest exchange coupling parameter $I_{44}(1)$=182.2~K couples magnetic
moments of Fe4 atoms with a distance $r_{44}(1)$=2.490~\AA.
Having at hand values of exchange interaction constants and local magnetic moments for
all Fe magnetic sublattices it is straightforward to estimate $T_C$ using
nearest neighbors mean-field approximation. For this purposes
absolute value of spin vector is $S_i=\mu_{Fe(i)}/2\mu_B$ ($i$=1,2,3,4).
To do that first one should define a set of self-consistent equations
for average value of $z$-component of the spin $m_i=<S_i^z>$, which are
\begin{equation}
m_i=S_i{\cal L}\Bigl(\frac{h_iS_i}{k_BT}\Bigr),
\label{eq24}
\end{equation}
where $h_i$ is Weiss field acting on spin $S_i$ from nearest neighbors.
Next one should linearize right hand sides of Eqs.~\ref{eq24}
expanding Langevin function for $T\rightarrow T_C$
\begin{equation}
m_i=\frac{S_i^2}{3}\frac{h_i}{k_BT_С}.
\label{eq25}
\end{equation}
Taking into account rhombohedral crystal structure of Gd$_2$Fe$_{17}$
one can derive set of linear equations for $m_i$ defining $T_C$ 
\begin{eqnarray}
\label{eq26}
m_1 = \beta_C \frac{S_1^2}{3} & [ & I_{11}(1)z_{11}(1)m_1 + I_{12}(1)z_{12}(1)m_2 + \nonumber \\
                            & + &  I_{13}(1)z_{13}(1)m_3 + I_{14}(1)z_{14}(1)m_4], \nonumber \\
m_2 = \beta_C \frac{S_2^2}{3} & [ & I_{21}(1)z_{21}(1)m_1 + I_{23}(1)z_{23}(1)m_2 + \nonumber \\
                            & + &  I_{33}(1)z_{33}(1)m_3], \nonumber \\
m_3 = \beta_C \frac{S_3^2}{3} & [ & I_{31}(1)z_{31}(1)m_1 + I_{32}(1)z_{32}(1)m_2 + \nonumber \\
                            & + & I_{33}(1)z_{33}(1)m_3+\nonumber \\
                            & + & (I_{34}(1)z_{34}(1) + I_{34}(2)z_{34}(2))m_4], \nonumber \\
m_4 = \beta_C \frac{S_4^2}{3} & [ & I_{41}(1)z_{41}(1)m_1 + I_{42}(1)z_{42}(1)m_2+ \nonumber \\
                            & + & (I_{43}(1)z_{43}(1) + I_{43}(1)z_{43}(1))m_3 + \nonumber \\
                            & + & I_{44}(1)z_{44}(1)m_4],
\end{eqnarray}
corresponding exchange integrals $I_{ij}(1)$ between Fe$i$ and Fe$j$ crystallographic classes in
the first coordination sphere
and nearest neighbors number $z_{ij}(1)$ on a distance $r_{ij}(1)$ are taken from Table~\ref{tab3}.
At that for cumbersome crystal structures  $z_{ij}(1)\neq z_{ji}(1)$ but $n_iz_{ij}(1)=n_jz_{ji}(1)$,
where $n_i$ number of atoms of sort Fe$i$ in a elementary magnetic cell.

As was obtained in Ref.~\onlinecite{Luk09} Eqs.~(\ref{eq26}) give $T_C$=429~K for rhombohedral Gd$_2$Fe$_{17}$
which is slightly smaller than experimental one 475~K\cite{Kou98}.
A reason of this discrepancy is mainly absence in our model of exchange interactions between Fe and Gd
sublattices and oscillating exchange interactions with next coordination spheres.
However this result shows that largest part (leading contribution) of $T_C$ comes from
exchange interactions of Fe sublattice between nearest neighbors. Thus it is interesting to
explore influence of strongest exchange $I_{11}(1)$ ``in dumbbell'' on $T_C$.

At the beginning hypothetically one switches off the interaction  $I_{11}(1)$.
Then solution of Eqs.~(\ref{eq26}) leads to $T_C(I_{11}(1)=0)$=414~K which is only
3.5\% less than in Ref.~\onlinecite{Luk09}. Now we consider opposite case
$I_{11}(1) \gg I_{ij}(1)$, i.e. consider dumbbell pair as dimer.
Thus one can apply approach described in Sec.~\ref{s2}:
Fe1-Fe1 cluster with strong exchange coupling $I_{11}(1)$ will be
treated exactly while other exchange bonds within mean-field approach.
After usual linearization of equations on $m_1=<S_1^z>$ for small values of
$m_i~(i=1,2,3,4)$ close to $T_C$ one gets equation
\begin{eqnarray}
m_1 &    =   & \beta_C \frac{S_1^2}{3} [1+{\cal L}(\beta_C I_{11}(1) S_1^2)]\times \\ \nonumber
    & \times & [I_{12}(1)z_{12}(1)m_2 + I_{13}(1)z_{13}(1)m_3 + \\ \nonumber
    &    +   &  I_{14}(1)z_{14}(1)m_4],
\label{eq27}
\end{eqnarray}
which should substitute the first equation of Eqs.~(\ref{eq26}).

For the rhombohedral phase of Gd$_2$Fe$_{17}$ second largest exchange
parameter is $I_{44}(1)$=182.2~K. Let us take $I_{11}(1)=10I_{44}(1)$=1822~K then
$T_C(I_{11}(1)=10I_{44}(1))$=456~K. And finally for the limit
 $I_{11}(1)\rightarrow\infty$, when ${\cal L}(\beta_C I_{11}(1) S_1^2)\to 1$,
Curie temperature is $T_C(I_{11}(1)\rightarrow\infty)$=464~K. 
Thereby even dumbbell exchange $I_{11}(1)\rightarrow\infty$ gives rise of $T_C$
only 10\%.
Also from Eqs.~(\ref{eq27}) one follows that for $I_{11}(1)\rightarrow\infty$
between spins $S_1$ and $S_2$ Weiss field acting from other spins
of atoms Fe2, Fe3 and Fe4 is doubled.

Another interesting observation one can do for
Curie point $T_C$ calculated from averaged over all types
of Fe atoms values of exchange constant $\bar I$, spin value $\bar S$ and
number of nearest neighbors $\bar z$
\begin{eqnarray}
k_BT_C &  =  & \frac{1}{3} \bar I (\bar S)^2 \bar z, \\ \nonumber
\bar S &  =  & \sum\limits_{i=1}^{4} n_i S_i / \sum\limits_{i=1}^{4} n_i = 1.12, \\ \nonumber
\bar z &  =  & \sum\limits_{i=1,j}^{4} n_i z_{ij} / \sum\limits_{i=1}^{4} n_i =10, \\ \nonumber
\bar I &  =  & 3 k_BT_C / (\bar S)^2 \bar z = 102.6~K,
\label{eq28}
\end{eqnarray}
here $T_C$=429~K.
From Table~\ref{tab3} one can see that most of values of exchange integrals are very close to the value of $\bar I$.
Thus these Fe-Fe exchange bonds are responsible for the Curie point value rather than 
Fe1-Fe1 exchange bond only, independently how strong it is.

\section{Conclusion}
\label{conclusion}

In this work we proposed description of several model magnetic structures
with diluted magnetic dimers which do not form
any infinite magnetic cluster. The selected strong
exchange bond $I_A$ is supposed to be much larger than other exchange interactions
$I_B$ in the elementary magnetic cell. Following ideas of Ref.~\onlinecite{Scal75}
the magnetic dimer is treated exactly while other couplings are treated
within mean-field approximation. In contrast to regular mean-field approximation
latter approach allows one to obtain finite
Curie temperatures $T_C$ even for $I_A\rightarrow\infty$ which is physically correct.
Also if was demonstrated that for $0\leq I_A/I_B < \infty$ $T_C$ deviates just about $\pm$10\%
from the value $T_C(I_A=I_B)$ and is getting closer to that while number of atoms
in the elementary magnetic cell grows.
For the case $I_A\rightarrow\infty$ we obtained doubling of spin value of spins forming a dimer,
which corresponds to doubling of Weiss field acting on the spin from other spins
in the elementary magnetic cell.
After all one can conclude that such diluted magnetic dimers do not influence very much
on $T_C$ value of the whole system in the case $I_A\rightarrow\infty$, and then even
less for finite $I_A$.

Based on these results we perform analogous calculations of $T_C$ for real system --
the rhombohedral phase of Gd$_2$Fe$_{17}$. There is so called dumbbell Fe1-Fe1 dimer
with the largest exchange interaction value $I_{11}(1)$ in the system.
We showed in this investigation that for such complicated
crystal structure $T_C$ is mainly defined by weaker exchange interactions of Fe1
with other Fe atoms in the elementary magnetic cell rather than by
the $I_{11}(1)$ exchange only.

\section{Acknowledgments} 
This work is partly supported by RFBR grant 08-02-00021 and was performed
within the framework of programs of fundamental research of the Russian Academy
of Sciences (RAS) ``Quantum physics of condensed matter'' (09-$\Pi$-2-1009) and 
of the Physics Division of RAS  ``Strongly correlated electrons in solid states'' 
(09-T-2-1011). IN thanks Grant of President of Russia MK-614.2009.2,
interdisciplinary UB-SB RAS project, and Russian Science Support
Foundation.






\end {document}